\documentstyle[preprint,pra,aps]{revtex}

\begin{document}


\tighten

\title{Photon echo signature of vibrational superposition states created by 
femtosecond excitation of molecules}

\author{Holger F. Hofmann, Takao Fuji, and Takayoshi Kobayashi}
\address{Department of Physics, Faculty of Science, University of Tokyo\\
7-3-1 Hongo, Bunkyo-ku, Tokyo113-0033, Japan}

\date{\today}

\maketitle

\begin{abstract}
A pair of coherent femtosecond pulse excitations applied to a 
molecule with strong electron-phonon coupling creates a coherent
superposition of a low momentum and a high momentum wavepacket
in the vibrational states of both the excited state and the ground state
of the coherent transition. As the excited state is accelerated further,
interference between the high momentum ground state contribution
and the low momentum excited state contribution causes a photon echo.
This photon echo is a direct consequence of quantum interference
between separate vibrational trajectories and can therefore provide 
experimental evidence of the non-classical properties of molecular 
vibrations.
\end{abstract}
\pacs{PACS numbers:
42.50.Ct  
42.50.Md  
33.80.-b  
42.65.Re  
}

``{\it The feature of quantum mechanics which most distinguishes it from
classical mechanics is the coherent superposition of distinct physical 
states}''\cite{Wal94}. In the case of molecular vibrations, the 
motion of a single Gaussian wavepacket usually corresponds well with the
classical motion for the potential considered. However, quantum mechanics 
also allows coherent superpositions between two wavepackets with distinct
positions and momenta. The properties of such superpositions depend on the
phase between the two Gaussian wavepackets, a property that has no analog 
in the classical theory. The study of such superpositions should therefore 
provide insights into the non-classical features of quantum mechanics.

The creation and measurement of vibrational wavepackets by optical 
excitations has been studied both experimentally and theoretically
\cite{Pei88,Kra93,Jan94,Koh95,Kis96,Shn96,Oht99,Sta00,Her00}. 
In particular, the creation of a coherent superposition of two distinct
Gaussian wavepackets (also referred to as a ``cat state'') in a molecular 
vibration has been predicted if the molecule is excited by a sequence of two 
femtosecond pulses \cite{Jan94}. However, it is difficult to obtain 
experimental evidence indicating the successful creation of the coherent 
superposition. In the following, it will be shown that the nonlinear optical 
contributions in the formation of the superposition state automatically 
produce a photon echo effect that corresponds to the ``which path'' 
interference between two distinct trajectories of the molecular vibration. 
Thus, the vibrational photon echo may provide direct experimental evidence 
of quantum coherence between two vibrational wavepackets.

The Hamiltonian describing the electron-phonon interaction of an electronic
two level system and a single vibrational degree of freedom may be written
as 
\begin{equation}
\label{eq:Htot}
\hat{H}_0 = \frac{\hat{p}^2}{2 m} + 
  V_G(\hat{x})\otimes \!\mid\! G \rangle \langle G \!\mid\!
+ V_E(\hat{x})\otimes \!\mid\! E \rangle \langle E \!\mid\!,
\end{equation}
where $m$ is the effective mass of the vibration, and $V_G(\hat{x})$ and
and $V_E(\hat{x})$ describe the vibrational potentials associated with the
electronic ground state $\mid \! G \rangle$ and the electronic excited state
$\mid \! E \rangle$, respectively. The position operator $\hat{x}$
and the conjugate momentum operator $\hat{p}$ represent the dynamical 
variables of the vibrational mode under consideration. 

Initially, the molecular system is in the electronic ground state 
$\mid\! G \rangle$ and the vibrational state $\mid \psi_0 \rangle$ is
localized near the minimum of the ground state potential $V_G(\hat{x})$.
It is therefore convenient to define this minimum as $x=0$ and its
potential as $V_G(0)=0$. If the timescales considered are much shorter 
than the period of a molecular vibration, the vibrational wavefunction 
will always remain close to $x=0$. Moreover, the momentum is also close
to zero initially, and its changes can be considered small enough to
neglect the quadratic term $\hat{p}^2/m$. The total Hamiltonian can 
then be linearized in $\hat{x}$ and $\hat{p}$. The approximate Hamiltonian 
reads
\begin{eqnarray}
\label{eq:Ham}
\hat{H}_0 &\approx& \left(\hbar\omega_0 - F_E\hat{x}\right)
\otimes \!\mid\! E \rangle \langle E \!\mid\!,
\nonumber \\[0.3cm] &&
\mbox{with}\hspace{0.5cm} 
\hbar \omega_0 = V_E(x=0)
\nonumber \\ &&
\mbox{and}\hspace{0.5cm}
F_E=-\frac{\partial}{\partial x} V_E(x)\big|_{x=0}.
\end{eqnarray}
This Hamiltonian describes the linear acceleration of the excited state 
component of the vibrational state by the force $F_E$. In terms of the 
momentum eigenstate components $\psi_{G}(p;t) = 
\langle G ; p \!\mid\! \psi(t)\rangle$ and $\psi_{E}(p;t) = 
\langle E ; p \!\mid\! \psi(t)\rangle$, this acceleration can be written as
\begin{eqnarray}
\psi_G(p;t) &=& \psi_G(p;0)
\nonumber \\
\psi_E(p;t) &=& \exp(-i\omega_0 t) \psi_E(p\!-\!F_E\,t;\; 0).
\end{eqnarray} 
Note that this evolution of the vibrational wavefunction preserves the
quantum coherence between the excited state and the ground state contributions.
It is therefore not possible to assign separate ``realities'' to 
excited state and ground state molecules. Instead, any coherent overlap
between the vibrational states corresponds to a coherent electronic dipole.
This coherent dipole is given by the operator 
$\hat{d}=\mid \! G\rangle\langle E \mid$. Its expectation value reads 
\begin{equation}
\langle \hat{d} \rangle (t) = \int d\! p \psi_G^*(p;t)\psi_E(p;t).
\end{equation}
The electronic dipole of the molecular transition thus represents an 
interference between the accelerated excited state and the non-accelerated
ground state. The acceleration process separates the vibrational state
like a beam splitter separates the incoming fields. In a photon echo 
experiment, the first pulse at $t_0-\tau$ splits the vibrational dynamics,
the second pulse at $t_0$ ``reflects'' the excited state into the ground
state and vice versa, and the photon echo indicates interference between
two indistinguishable paths of acceleration. The ground state component 
of the photon echo dipole corresponds to acceleration during 
$t_0-\tau<t<t_0$ followed by a constant momentum of $F_E \tau$ during 
$t_0<t<t_0+\tau$, and the excited state component corresponds to zero momentum
during $t_0-\tau<t<t_0$ followed by acceleration to a momentum of $F_E \tau$  
during $t_0<t<t_0+\tau$. These trajectories are illustrated in figure 
\ref{trajectories}. 

In the following, we apply this description of the molecular dynamics to 
a pair of ultrafast excitations at times $t_0-\tau$ and $t_0$. The pulses
are considered to be much shorter than $\tau$. Before the first pulse,
the molecule is in its ground state, given by $\psi_G(p)=\psi_0(p)$ and
$\psi_E(p)=0$. Between the two pulses ($t_0-\tau<t<t_0$), the coherent
evolution of the partially excited state is given by
\begin{eqnarray}
\psi_G(p;t) &=& \cos(\phi/2) \psi_0(p) 
\nonumber \\
\psi_E(p;t) &=& 
e^{-i\omega_0 (t-t_0+\tau)}
\sin(\phi/2)\psi_0\left(p\!-\!F_E(t\!-\!t_0+\tau)\right),
\end{eqnarray}
where $\phi$ is a measure of the pulse area exciting the molecule.
The expectation value of the coherent dipole evolves according to
\begin{equation}
\langle\hat{d}\rangle(t) = e^{-i\omega_0 (t-t_0+\tau)}
\frac{1}{2}\sin(\phi) \int d\!p \psi_0^*(p)
\psi_0(p\!-\!F_E(t\!-\!t_0+\tau)),
\end{equation}
which corresponds to the autocorrelation of the vibrational wavefunction 
in momentum space, $\psi_0(p)$. 
The dipole dephasing time $t_\phi$ is therefore given by 
\begin{equation}
t_\phi=\frac{\delta\!p}{F_E},
\end{equation}
where $\delta\!p$ is the momentum uncertainty of the initial wavepacket.
If $\delta\!p$ is much smaller than $F_E \tau$, then the coherent dipole 
will be close to zero at $t=t_0$.
 
The second pulse at $t=t_0$ then restores dipole coherence by transferring 
part of the ground state component to the excited state and vice versa.
The evolution of the total molecular state for $t>t_0$ reads
\begin{eqnarray}
\psi_G(p;t) &=& \frac{1}{2}\left(1+\cos(\phi)\right) \psi_0(p) 
              - e^{-i\omega_0 \tau}
                \frac{1}{2}\left(1-\cos(\phi)\right) 
                            \psi_0\left(p\!-\!F_E\tau\right)
\nonumber \\
\psi_E(p;t) &=& e^{-i\omega_0(t-t_0)} \frac{1}{2}\sin(\phi)
        \psi_0\left(p\!-\!F_E(t\!-\!t_0)\right)
\nonumber \\ &&
   + e^{-i\omega_0(t-t_0+\tau)} \frac{1}{2}\sin(\phi)
         \psi_0\left(p\!-\!F_E\tau\!-\!F_E(t\!-\!t_0)\right).
\end{eqnarray}
The total vibrational state now consists of four separate contributions.
Initially ($t=t_0$), there is dipole coherence between two pairs, the one 
around $p=0$ and the one around $p=F_E\tau$. This coherence is 
lost as the excited state is accelerated. The dipole dynamics of the 
decoherence process reads
\begin{equation}
\langle\hat{d}\rangle(t) = e^{-i\omega_0 (t-t_0)}
\frac{1}{2}\sin(\phi)\cos(\phi) \int d\!p \psi_0^*(p)
\psi_0(p\!-\!F_E(t\!-\!t_0)).
\end{equation}
This corresponds to the linear response of the partially excited two level
system to the second pulse.
However, there is a revival of
the dipole coherence in the form of a photon echo as the excited state 
from $p=0$ is accelerated to $p=F_E\tau$ and interferes with the ground state
component there. Figure \ref{wavepackets} illustrates the coherent 
wavefunction at $t=t_0+\tau$. The dipole dynamics close to $t=t_0+\tau$ 
are given by 
\begin{equation}
\langle\hat{d}\rangle(t) = e^{-i\omega_0 (t-t_0-\tau)}
\frac{1}{4}\sin(\phi)\left(1-\cos(\phi)\right)
 \int d\!p \psi_0^*(p\!-\!F_E\tau)
\psi_0(p\!-\!F_E(t\!-\!t_0)).
\end{equation}
This result is again equal to the autocorrelation of $\psi_0(p)$, but it
is centered around $t=t_0+\tau$. Figure \ref{pulses} shows the sequence of 
pulses and the dipole response. Since the first two dipole signals are
immediate responses to the exciting pulses, they suddenly appear at $t_0-\tau$
and at $t_0$, followed by a gradual decay given by the autocorrelation of 
$\psi_0(p)$. The echo pulse arises from the hidden coherence in the dynamics 
following the second pulse at $t=t_0$. It therefore appears gradually and
is symmetric around $t=t_0$. 
Since dipole coherence always indicates an interference between ground
state and excited state components, the echo indicates equal momentum of the 
accelerated excited state and the non-accelerated ground state. In terms of
the average momentum of the four separate contributions, figure 
\ref{trajectories} shows the trajectories involved in the quantum 
interferences indicated by
the dipole expectation value $\langle \hat{d} \rangle$. By comparing the 
dipole evolution given in figure \ref{pulses} with the trajectories in
figure \ref{trajectories}, the optical signals can be related to the
quantum dynamics of the vibration.

The analogy between photon echoes in inhomogeneously broadened transitions
and the molecular photon echoes discussed here arises from the assumption 
that the position coordinate $\hat{x}$ remains nearly constant during the 
experiment. The inhomogeneity is then a consequence of the randomness of the
position coordinate $\hat{x}$ given by the spatial width of the vibrational 
wavepacket. The main difference between the photon echo in an inhomogeneously
broadened medium and the vibrational photon echo discussed here is that the
coherence of the contributions from different positions corresponds to a
well defined momentum. It is therefore impossible to identify each precise
position $x$ with a different molecule, since this would imply an infinite 
momentum uncertainty. For most practical purposes, however, the experimental
setup corresponds to a conventional photon echo experiment. 

In order to satisfy the assumption that $\hat{x}$ does not change
much during the experiment, the spatial shifts induced by the velocity $p(t)/m$
during the delay time $\tau$ must be much 
smaller than the spatial width of the vibrational wavepacket.
The shift in position can be determined by integrating the velocity
$p(t)/m$ over time. The total shift of position thus contains a ``memory'' 
of the momentum path taken by the molecular vibration, destroying the quantum
interference. At the interference point, the difference in position between
the ground state component and the excited state component with $p=F_E\tau$
is
\begin{equation}
\Delta x = \frac{F_E \tau^2}{m}.
\end{equation} 
In momentum representation, this shift appears as a phase factor of
$\exp(-ip\Delta x/\hbar)$. The maximal coherent dipole of the echo pulse 
is then reduced by a factor of
\begin{equation}
\xi = \int d\!p \exp\left(-i\frac{\Delta x}{\hbar}p\right)|\psi_0(p\!)|^2. 
\end{equation}
Since the ground state of a harmonic oscillator of frequency $\Omega$ is
a Gaussian wavepacket with a momentum variance of 
$\delta\!p^2 = \hbar \Omega m/2$, the decoherence factor $\xi$ can be 
written as
\begin{eqnarray}
\label{eq:deco}
\xi &=& \exp\left(-\frac{\delta\!p^2 \Delta x^2}{2 \hbar^2} \right) 
\nonumber \\ &=&
\exp\left(-\frac{F_E^2\Omega}{4 \hbar m} \tau^4 \right).
\end{eqnarray}
The dependence of the decoherence factor $\xi$ on delay time $\tau$ is
shown in figure \ref{deph}.
This fourth order exponential decay should be a typical signature of
molecular photon echoes, allowing a distinction between such echoes and
the echo effect caused by inhomogeneous broadening.
The decoherence time $T$ is given by 
\begin{eqnarray}
\label{eq:T}
T= \left(\frac{4 \hbar m}{F_E^2\Omega}\right)^{\frac{1}{4}}.
\end{eqnarray} 
An estimate of the dephasing time can be obtained from typical 
values of $F_E\approx 10^{-8} N$, $\Omega \approx 10^{14}s^{-1}$,  
and $m \approx 10^{-25} kg$. For these values, the dephasing time is
about $10^{-14} s$ or ten femtoseconds. In order to obtain a clear 
separation between the two pulses exciting the molecule,
it is therefore necessary to use extremely short pulses. 
If the pulses do overlap, an echo can still be obtained, but 
since the vibrational wavepackets overlap as well the interference 
cannot be traced to two separate trajectories. 

Equation (\ref{eq:T}) suggests that optimal results can be obtained in a 
molecular vibration with a high effective mass $m$, a low vibrational 
frequency $\Omega$ in the ground state, and a small force $F_E$ on the excited
state. However, a strong force $F_E$ is necessary to ensure a short dipole
dephasing time. Using $\delta\!p^2 = \hbar \Omega m/2$, the ratio of 
decoherence time $T$ and dephasing time $t_\phi$ is
\begin{equation}
\frac{T}{t_\phi} = 
2 \left( \frac{F_E^2}{\hbar m \Omega^3}\right)^{\frac{1}{4}}.
\end{equation}
For the typical values given above, this ratio is about three. 
In order to obtain a higher ratio, it is desirable to have a lower 
effective mass $m$, a lower vibrational frequency $\Omega$, and a 
stronger force $F_E$.
The best system for observing a vibrational photon echo would thus be a
system with an unusually low ground state vibrational frequency. Possibly,
a very unstable bond with a shallow low curvature potential could be used.
A transition to an anti-bonding state could then trigger a dissociation 
process with a sufficiently high initial accelerating force $F_E$ to provide
rapid dephasing. 

In conclusion, the results presented here highlight the possibility of using 
nonlinear femtosecond spectroscopy to probe the quantum nature of vibrational 
states. It also reveals fundamental quantum mechanical 
details of the interaction between the coherent dipole dynamics and the 
vibrational dynamics in molecular systems, thus providing some insights into 
the role of vibrational quantum coherence in the nonlinear optics of molecules.
This method could thus provide a starting point for a more general 
investigation of interference effects in the dynamics of bound atoms.

\section*{Acknowledgements}
One of us (HFH) would like to acknowledge support from the Japanese 
Society for the Promotion of Science, JSPS.



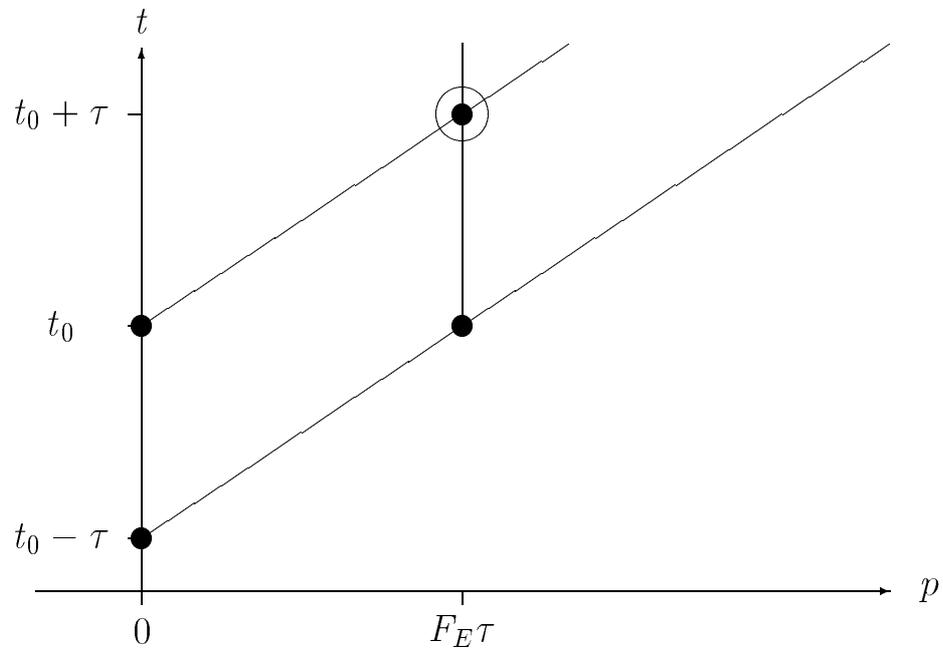
\begin{figure}
\caption{\label{trajectories}
Schematic representation of the momentum trajectories describing
the acceleration of the molecular vibration in response to the 
femtosecond excitations at $t_0-\tau$ and $t_0$. Filled circles mark
quantum interferences between the ground and excited state trajectories. 
The circles at $t_0-\tau$ and $t_0$ correspond to the dipole response 
caused by the short pulse excitations. The double circle at $t_0+\tau$ 
does not coincide with the excitations and therefore marks the coherent 
dipole of the photon echo.}
\end{figure}
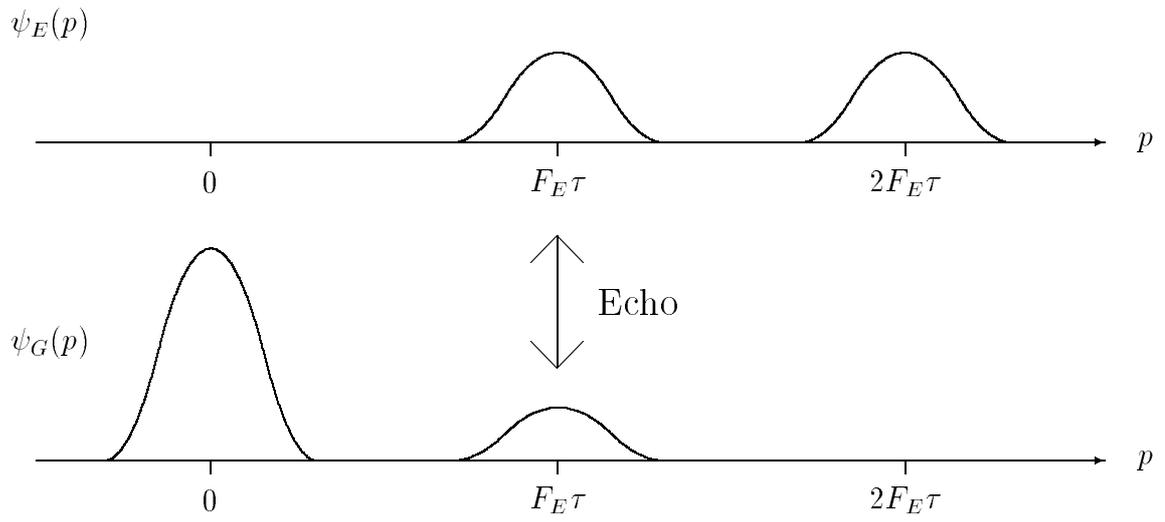
\begin{figure}
\caption{\label{wavepackets} Qualitative illustration of the vibrational 
wavefunction in momentum space representation at $t_0+\tau$ for a pair of 
exciting pulses 
with a pulse area of $\phi=\pi/3$ each. The coherent dipole 
oscillation observed as photon echo originates from the quantum 
interference of the vibrational cat state components at a momentum 
of $F_E\tau$. }
\end{figure}
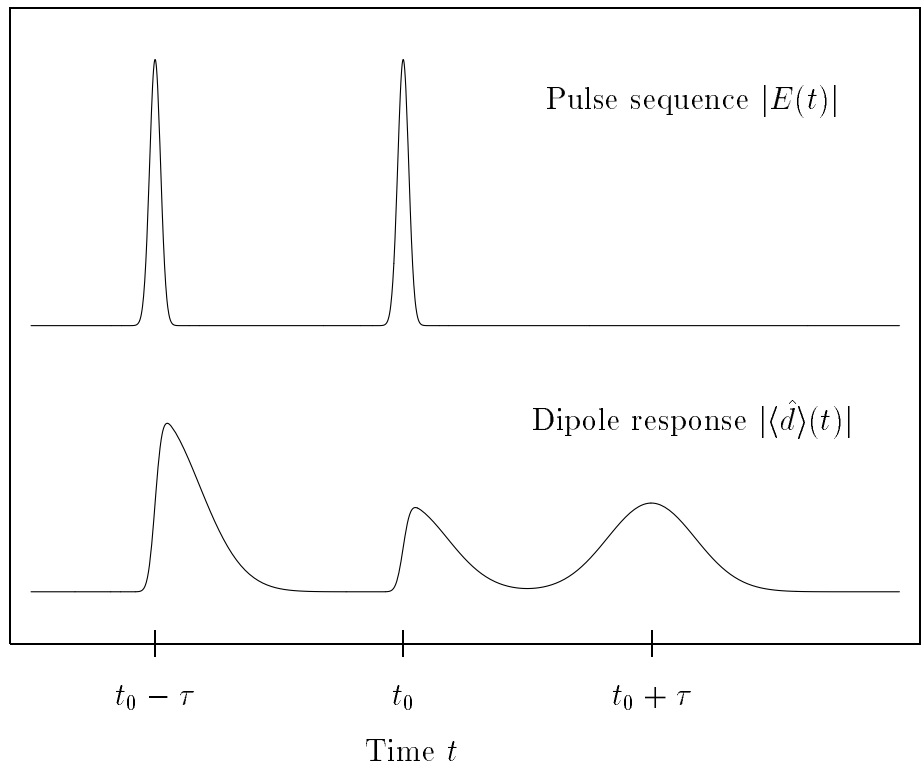
\begin{figure}
\caption{\label{pulses} Qualitative illustration of the pulse sequence 
$E(t)$ (top) and the dipole response $|\langle\hat{d}\rangle(t)|$
(bottom) for $\phi=\pi/3$. Note that the echo pulse is symmetric 
around $t=t_0+\tau$.}
\end{figure}
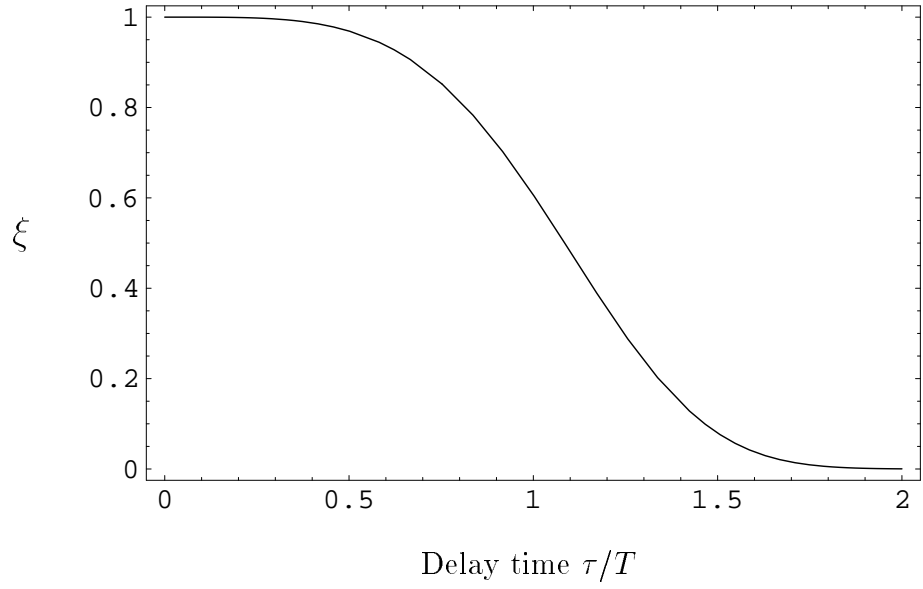
\begin{figure}
\caption{\label{deph} Decoherence factor $\xi$ as a function of delay
time $\tau$. $T$ is the decoherence time of the vibrational photon echo.}
\end{figure}
\end{document}